\documentclass[aps,preprint,showpacs,groupedaddress]{revtex4-1}
\usepackage{amssymb}
\usepackage{mathrsfs}
\usepackage{amsmath}
\usepackage{mathtools}
\usepackage{pstricks}
\usepackage{color}
\usepackage{graphicx}
\usepackage{amsthm}

\usepackage{physics}

\begin{document}


\title{\bf Exact Lorentz-violating $q$-deformed O($N$) universality class}



\author{P. R. S. Carvalho}
\email{prscarvalho@ufpi.edu.br}
\affiliation{\it Departamento de F\'\i sica, Universidade Federal do Piau\'\i, 64049-550, Teresina, PI, Brazil}

\author{M. I. Sena-Junior}
\email{marconesena@poli.br}
\affiliation{\it Escola Polit\'{e}cnica de Pernambuco, Universidade de Pernambuco, 50720-001, Recife, PE, Brazil}




\begin{abstract}
We examine the influence of exact Lorentz-violating symmetry mechanism on the radiative quantum corrections to the critical exponents for massless $q$-deformed O($N$) $\lambda\phi^{4}$ scalar field theories. For that, we employ three different and independent field-theoretic renormalization group methods for computing analytically the $q$-deformed critical exponents up to next-to-leading order. Then we generalize the former finite loop level results for any loop order. We show that the Lorentz-violating $q$-deformed critical exponents, obtained through the three methods, turn out to be identical and furthermore the same as their Lorentz-invariant $q$-deformed ones. We argue that this result is in accordance with the universality hypothesis. 
\end{abstract}


\maketitle


\section{Introduction} 

\par It is known nowadays that there is a relation between a $q$-deformation and a cosmological constant. In fact, this deformation can produce a cosmological constant term in $3d$ quantum gravity \cite{Livine2017} as well as in others problems involving gravity theories \cite{Dil,1751-8121-42-42-425402,0264-9381-24-13-009,MAJOR1996267}. On the other hand, the study of $q$-deformation has not been constrained only to the gravity scenario but has also attracted great attention in distinct physics research branches in the last few years \cite{PhysRevLett.106.241602,AGANAGIC2005304,PhysRevE.61.1218,Delduc2014,PhysRevA.63.023802,0305-4470-35-43-316,CONROY20104581,GRIGUOLO20071,BASEILHAC2006309,1126-6708-2006-01-035,Adv.High.EnergyPhys.20179530874,Eur.Phys.J.Plus1322017398,Int.J.Theor.Phys.5620171746,Phys.DarkUniv.1620171,Adv.HighEnergyPhys.20179371391,EPL11320162000,JHEP1220160630,Phys.Rev.D912015044024,G.Vinod}. Particularly, in a recent paper \cite{PhysRevD.97.105006}, the radiative quantum corrections to the critical exponents for $q$-deformed O($N$) $\lambda\phi^{4}$ scalar field theory were computed up to next-to-leading-order (NLO). The set of these $q$-deformed critical exponents defines the $q$-deformed O($N$) universality class. A given universality class is characterized by many distinct physical systems undergoing a continuous phase transition whose scaling critical behaviors are described by the same set of (universal) critical exponents. This happens only if the many physical systems share some universal parameters as their dimension $d$, $N$ and symmetry of some $N$-component order parameter if the interactions of their constituents are of short- or long-range type. This is in essence the content of the universality hypothesis. On the other hand, the critical exponents do not depend on nonuniversal parameters as the critical temperature or the form of the lattice \cite{Stanley}. In this work we have to compute the effect of a symmetry breaking mechanism on the $q$-deformed critical exponents values, namely the Lorentz one \cite{EDWARDS2018319,Borges.Ferrari.Barone,PhysRevD.98.035018,PhysRevD.97.055048,Int.J.Geom.MethodsMod.Phys.13.1650049,doi:10.1142/S021988781850086X,BaetaScarpelli2017,PhysRevD.96.045019,doi:10.1142/S0217732318501158,PhysRevD.95.056005,doi:10.1142/S0217732315500376,Carvalho2013850,Carvalho2014320,PhysRevD.87.045012,PhysRevD.84.065030,PhysRevD.73.065015,ALTSCHUL2006679,PhysRevD.70.016001,PhysRevD.65.091701,universe2201630,PhysRevD.40.1886,PhysRevD.39.683,Kostelecký1991545,PhysRevD.59.124021,Amelino-Camelia20135,PhysRevLett.90.211601,Hayakawa200039,PhysRevLett.87.141601,PhysRevD.92.045016,doi:10.1142/S0217751X15500724,VANTILBURG2015236,PhysRevD.86.125015,PhysRevD.84.065030,Carvalho2013850,Carvalho2014320}. We have to treat this mechanism exactly in the Lorentz-violating parameters $K_{\mu\nu}$ \cite{PhysRevD.58.116002}. In fact, the effect of Lorentz violation has been probed recently in the critical exponents values for nondeformed O($N$) $\lambda\phi^{4}$ scalar field theories \cite{CARVALHO2017290,Carvalho2017} considering $K_{\mu\nu}$ exactly.   

\par In this paper we have to employ field-theoretic renormalization group and $\epsilon$-expansion techniques for renormalizing a massless $q$-deformed O($N$) $\lambda\phi^{4}$ scalar field theory. The renormalization program is a tool and was designed to get rid the divergences of a initially divergent field theory \cite{Wilson197475}. These divergences are a result of the $q$-deformed fields when evaluated at the same point of spacetime and expressed as the commutation relations of the $q$-deformed quantum fields. In the present case, this $q$-deformed quantum field is given by
\begin{eqnarray}\label{ygfdxzsze}
\phi_{q}(x) = \int \frac{d^{3}k}{(2\pi)^{3/2}2\omega_{k}^{1/2}}[a(k)_{q}\exp^{-ikx} +a_{q}^{\dagger}(k)\exp^{ikx}]\quad
\end{eqnarray}
where $\omega_{k}^{2} = \vec{k}^{2} + m^{2}$ and its creation and destruction operators obey to the $q$-deformed commutation relations
\begin{eqnarray}\label{kljkkjij}
a(k)_{q}a_{q}^{\dagger}(k^{\prime}) - q^{-1}a_{q}^{\dagger}(k^{\prime})a(k)_{q} \equiv [a(k)_{q}, a_{q}^{\dagger}(k^{\prime})] = q^{N(k)}\delta(k - k^{\prime}), 
\end{eqnarray}  
\begin{eqnarray}\label{sfcvcgf}
[a(k)_{q}, a_{q}(k^{\prime})] = 0 = [a^{\dagger}(k)_{q}, a_{q}^{\dagger}(k^{\prime})], 
\end{eqnarray}    
where $N(k) = a_{q}^{\dagger}(k)a_{q}(k)$ is the $q$-deformed number operator. From the ensemble mean values for these $q$-deformed quantum fields, we obtain the correlation functions or equivalently the $1$PI vertex parts $\Gamma^{(N)}$. For the theory approached here it is sufficient to renormalize the primitively ones, namely the $\Gamma^{(2)}$, $\Gamma^{(4)}$ and $\Gamma^{(2,1)}$ \cite{Amit}. The divergences in a massless theory are in the infrared limit and they can be absorbed in three distinct and independent methods to be displayed below. The first one is the normalization conditions method \cite{Amit}, where the external momenta of Feynman diagrams are held at fixed values. The second of them is the minimal subtractions scheme \cite{Amit}. This method is general and more elegant than the earlier since the external momenta values are keep at arbitrary values. The third and last of the methods is the Bogoliubov-Parasyuk-Hepp-Zimmermann (BPHZ) one \cite{BrezinLeGuillouZinnJustin,Amit}, where now the theory is renormalized through the introduction of counterterm diagrams besides also leaving the external momenta of diagrams arbitrary. The critical exponents are computed from the critical scaling properties of the renormalized primitively $1$PI vertex parts. At the critical point, the primitively $1$PI vertex parts present some anomalous critical behavior and acquire anomalous dimensions. When these anomalous dimensions are computed at the nontrivial fixed point we obtain the critical exponents. The nontrivial fixed point is evaluated from the nontrivial solutions to the equation $\beta_{q} = 0$ for the $\beta_{q}$ function \cite{Amit}. As there are four scaling relations among the six critical exponents, we must compute only two of them independently, for example $\eta_{q}$ and $\nu_{q}$. As the critical exponents are universal quantities, they must be the same if computed through any of the renormalization schemes. The free propagators are massless, since the mass in this field-theoretic formulation, the mass of the $q$-deformed quantum field is represented by $m^{2} \propto T - T_{c}$, where $T$ is some arbitrary temperature and $T_{c}$ is the critical one. Thus the $q$-deformed free propagator of the theory in momentum space is given by $G_{0}(k) = \parbox{12mm}{\includegraphics[scale=1.0]{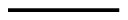}} = q/(k^{2} + K_{\mu\nu}k^{\mu}k^{\nu})$.    

\par This work will proceed as follows: Firstly we renormalize the theory through three distinct and independent field-theoretic renormalization group methods, namely the normalization conditions, minimal subtraction scheme and BPHZ ones, respectively. For each method we present the analytically evaluated needed Feynman diagrams and the corresponding $q$-deformed $\beta_{q}$ functions, anomalous dimensions and nontrivial fixed points. After that we compute the referred Lorentz-violating $q$-deformed critical exponents values at NLO for then at any loop level. At the conclusions we point out our final considerations.   

\section{Renormalization group methods and NLO Lorentz-violating $q$-deformed critical exponents}

\par Now we display the Feynman diagrams needed in the three renormalization methods as well as the corresponding $\beta_{q}$ functions, anomalous dimensions and nontrivial fixed points. The bare $1$PI vertex parts to be renormalized up to NLO are given by     

\begin{eqnarray}\label{gtfrdrdes}
\Gamma^{(2)}_{B} = \quad \parbox{12mm}{\includegraphics[scale=1.0]{fig9.eps}}^{-1} + \frac{1}{6}\hspace{1mm}\parbox{12mm}{\includegraphics[scale=1.0]{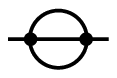}} + \frac{1}{4}\hspace{1mm}\parbox{10mm}{\includegraphics[scale=0.8]{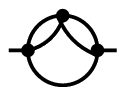}},
\end{eqnarray}
\begin{eqnarray}
\Gamma^{(4)}_{B} = \quad \parbox{12mm}{\includegraphics[scale=0.09]{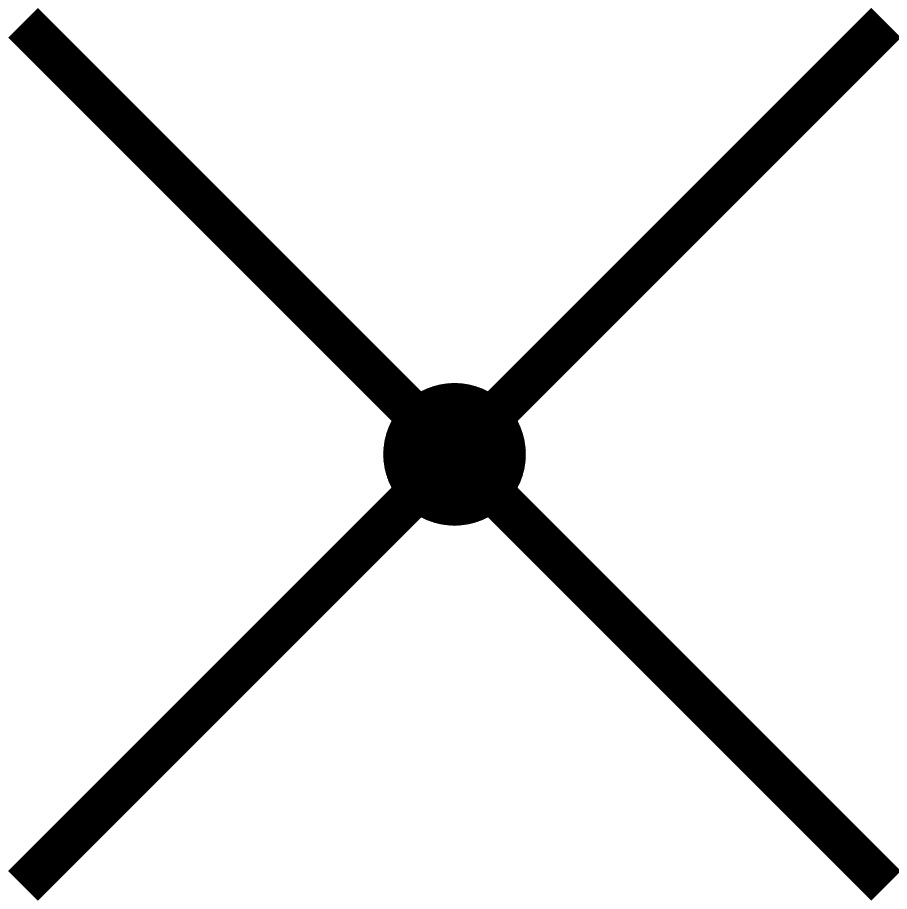}} +  \frac{1}{2}\hspace{1mm}\parbox{10mm}{\includegraphics[scale=1.0]{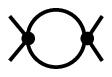}} + 2 \hspace{1mm} perm.  +  \frac{1}{4}\hspace{1mm}\parbox{16mm}{\includegraphics[scale=1.0]{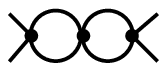}} + 2 \hspace{1mm} perm.  + \frac{1}{2}\hspace{1mm}\parbox{12mm}{\includegraphics[scale=0.8]{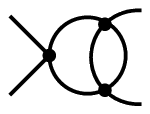}} + 5 \hspace{1mm} perm., 
\end{eqnarray}
\begin{eqnarray}\label{gtfrdesuuji}
&& \Gamma^{(2,1)}_{B} = \quad \parbox{14mm}{\includegraphics[scale=1.0]{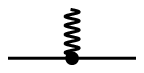}} \quad + \quad \frac{1}{2}\hspace{1mm}\parbox{14mm}{\includegraphics[scale=1.0]{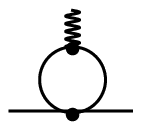}} +   \frac{1}{4}\hspace{1mm}\parbox{12mm}{\includegraphics[scale=1.0]{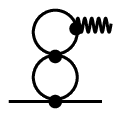}}  +  \frac{1}{2}\hspace{1mm}\parbox{12mm}{\includegraphics[scale=0.8]{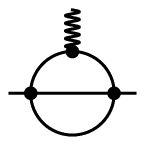}}.
\end{eqnarray}
While in the next two methods, we start from the bare theory to attain the renormalized one, in the last one we start from the renormalized theory just from the very beginning. 

\subsection{Normalization conditions}\label{Normalization conditions} 

\par In this method \cite{Amit,BrezinLeGuillouZinnJustin} the needed Feynman diagrams are computed at fixed external momenta values  
\begin{eqnarray}
\parbox{10mm}{\includegraphics[scale=1.0]{fig10.eps}}_{SP} = \frac{1}{\epsilon}\left(1 + \frac{1}{2}\epsilon  \right)q^{2}\mathbf{\Pi}, 
\end{eqnarray}   
\begin{eqnarray}
\parbox{12mm}{\includegraphics[scale=1.0]{fig6.eps}}^{\prime} = -\frac{1}{8\epsilon}\left( 1 + \frac{5}{4}\epsilon \right)q^{3}\mathbf{\Pi}^{2},
\end{eqnarray}  
\begin{eqnarray}
\parbox{10mm}{\includegraphics[scale=0.9]{fig7.eps}}^{\prime} = -\frac{1}{6\epsilon^{2}}\left( 1 + 2\epsilon  \right)q^{5}\mathbf{\Pi}^{3},
\end{eqnarray}  
\begin{eqnarray}
\parbox{12mm}{\includegraphics[scale=0.8]{fig21.eps}}_{SP} = \frac{1}{2\epsilon^{2}}\left( 1 + \frac{3}{2}\epsilon \right)q^{4}\mathbf{\Pi}^{2},
\end{eqnarray} 
where
\begin{eqnarray}
\parbox{10mm}{\includegraphics[scale=1.0]{fig10.eps}}_{SP} \equiv \parbox{10mm}{\includegraphics[scale=1.0]{fig10.eps}}\Bigg\vert_{P^{2}=1},
\end{eqnarray}
\begin{eqnarray}
\parbox{10mm}{\includegraphics[scale=1.0]{fig6.eps}}^{\prime} \equiv \frac{\partial}{\partial P^{2}}\parbox{11mm}{\includegraphics[scale=1.0]{fig6.eps}} \Bigg\vert_{P^{2}=1},
\end{eqnarray}
\begin{eqnarray}
\parbox{12mm}{\includegraphics[scale=.8]{fig21.eps}}_{SP} \equiv \parbox{12mm}{\includegraphics[scale=0.8]{fig21.eps}}\Bigg\vert_{P^{2}=1},
\end{eqnarray}
\begin{eqnarray}
\parbox{10mm}{\includegraphics[scale=1.0]{fig7.eps}}^{\prime} \equiv \frac{\partial}{\partial P^{2}}\parbox{12mm}{\includegraphics[scale=1.0]{fig7.eps}} \Bigg\vert_{P^{2}=1}
\end{eqnarray}
$\epsilon = 4 - d$ and the external momenta $P$ are written in terms of some arbitrary momentum scale $\kappa$ unit. The factor $\mathbf{\Pi} = 1/\sqrt{det(\mathbb{I} + \mathbb{K})}$ is the exact Lorentz-violating full factor \cite{PhysRevD.96.116002}. Thus the Lorentz-violating $\beta_{q}$ function, anomalous dimensions and nontrivial fixed point are given by
\begin{eqnarray}\label{jkjhigyg}
\beta_{q}(u) = -\epsilon u +   \frac{N + 8}{6}\left( 1 + \frac{1}{2}\epsilon \right)q^{2}\mathbf{\Pi}u^{2} -   \frac{3N + 14}{12}q^{4}\mathbf{\Pi}^{2}u^{3} + \frac{N + 2}{36}q^{3}(1-q)\mathbf{\Pi}^{2}u^{3}, 
\end{eqnarray}
\begin{eqnarray}
\gamma_{\phi ,q} = \frac{N + 2}{72}\left( 1 + \frac{5}{4}\epsilon \right)q^{3}\mathbf{\Pi}^{2}u^{2} - \frac{(N + 2)(N + 8)}{864}q^{5}\mathbf{\Pi}^{3}u^{3},\quad  
\end{eqnarray}
\begin{eqnarray}\label{sfsvcfdt}
\overline{\gamma}_{\phi^{2}, q}(u) = \frac{N + 2}{6}\left( 1 + \frac{1}{2}\epsilon \right)q^{2}\mathbf{\Pi}u -  \frac{N + 2}{12}q^{4}\mathbf{\Pi}^{2}u^{2},
\end{eqnarray}
where $\overline{\gamma}_{\phi^{2}}(u) = \gamma_{\phi^{2}}(u) - \gamma_{\phi}(u)$,
\begin{eqnarray}
u_{q}^{*} = \frac{6\epsilon}{(N + 8)q^{2}\mathbf{\Pi}}\Bigg\{ 1 +  \epsilon\Bigg[ \frac{3(3N + 14)}{(N + 8)^{2}} -\frac{1}{2} -\frac{N + 2}{(N + 8)^{2}}\frac{1 - q}{q} \Bigg]\Bigg\}.
\end{eqnarray}

\subsection{Minimal subtraction scheme} 

\par This method is characterized by its generality and elegance \cite{Amit,BrezinLeGuillouZinnJustin} since the Feynman diagrams are evaluated by keeping their external momenta at arbitrary values. Thus we obtain 
\begin{eqnarray}
\parbox{10mm}{\includegraphics[scale=1.0]{fig10.eps}} = \frac{1}{\epsilon} \left[1 - \frac{1}{2}\epsilon - \frac{1}{2}\epsilon L(P^{2} + K_{\mu\nu}P^{\mu}P^{\nu}) \right]q^{2}\mathbf{\Pi},\quad
\end{eqnarray}   
\begin{eqnarray}
\parbox{12mm}{\includegraphics[scale=1.0]{fig6.eps}} =  -\frac{P^{2} + K_{\mu\nu}P^{\mu}P^{\nu}}{8\epsilon}\left[ 1 + \frac{1}{4}\epsilon -2\epsilon L_{3}(P^{2} + K_{\mu\nu}P^{\mu}P^{\nu}) \right] q^{3}\mathbf{\Pi}^{2},
\end{eqnarray}  
\begin{eqnarray}
\parbox{10mm}{\includegraphics[scale=0.9]{fig7.eps}} =  -\frac{P^{2} + K_{\mu\nu}P^{\mu}P^{\nu}}{6\epsilon^{2}}\left[ 1 + \frac{1}{2}\epsilon -3\epsilon L_{3}(P^{2} + K_{\mu\nu}P^{\mu}P^{\nu}) \right] q^{5}\mathbf{\Pi}^{3},
\end{eqnarray}  
\begin{eqnarray}
\parbox{12mm}{\includegraphics[scale=0.8]{fig21.eps}} = \frac{1}{2\epsilon^{2}}\left[1 - \frac{1}{2}\epsilon - \epsilon L(P^{2} + K_{\mu\nu}P^{\mu}P^{\nu}) \right]q^{4}\mathbf{\Pi}^{2},
\end{eqnarray}
where $P$ are the external momenta and
\begin{eqnarray}\label{uhduhguh}
L(P^{2} + K_{\mu\nu}P^{\mu}P^{\nu}) = \int_{0}^{1}dx\ln[x(1-x)(P^{2} + K_{\mu\nu}P^{\mu}P^{\nu})],
\end{eqnarray}
\begin{eqnarray}\label{uhduhguhf}
L_{3}(P^{2} + K_{\mu\nu}P^{\mu}P^{\nu}) = \int_{0}^{1}dx(1-x)\ln[x(1-x)(P^{2} + K_{\mu\nu}P^{\mu}P^{\nu})].
\end{eqnarray}
The corresponding Lorentz-violating $\beta_{q}$ function, anomalous dimensions and nontrivial fixed point obtained are the ones
\begin{eqnarray}\label{rygyfggffgyt}
\beta_{q}(u) = -\epsilon u +   \frac{N + 8}{6}q^{2}\mathbf{\Pi}u^{2} - \frac{3N + 14}{12}q^{4}\mathbf{\Pi}^{2}u^{3} +  \frac{N + 2}{36}q^{3}(1-q)\mathbf{\Pi}^{2}u^{3}, 
\end{eqnarray}
\begin{eqnarray}\label{hkfusdrs}
\gamma_{\phi ,q} = \frac{N + 2}{72}q^{3}\mathbf{\Pi}^{2}u^{2} - \frac{(N + 2)(N + 8)}{1728}q^{5}\mathbf{\Pi}^{3}u^{3},  
\end{eqnarray}
\begin{eqnarray}\label{kujyhghsghju}
\overline{\gamma}_{\phi^{2}, q}(u) = \frac{N + 2}{6}q^{2}\mathbf{\Pi}u -  \frac{N + 2}{12}q^{4}\mathbf{\Pi}^{2}u^{2},
\end{eqnarray}
\begin{eqnarray}
u_{q}^{*} = \frac{6\epsilon}{(N + 8)q^{2}\mathbf{\Pi}}\Bigg\{ 1 +  \epsilon\Bigg[ \frac{3(3N + 14)}{(N + 8)^{2}} -\frac{N + 2}{(N + 8)^{2}}\frac{1 - q}{q} \Bigg]\Bigg\}.
\end{eqnarray}

\subsection{Massless Bogoliubov-Parasyuk-Hepp-Zimmermann method} 

\par In this method, as opposed to the earlier ones, we start from the renormalized theory 
\begin{eqnarray}\label{fdhjdii}
\Gamma^{(2)} = \parbox{12mm}{\includegraphics[scale=1.0]{fig9.eps}}^{-1}  - \frac{1}{6}\hspace{1mm}\parbox{12mm}{\includegraphics[scale=1.0]{fig6.eps}}  -  \frac{1}{4}\hspace{1mm}\parbox{10mm}{\includegraphics[scale=0.8]{fig7.eps}} - \frac{1}{3} \mathcal{K}
  \left(\parbox{10mm}{\includegraphics[scale=1.0]{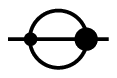}} \right), 
\end{eqnarray}
\begin{eqnarray}\label{fdijgihu}
&&\Gamma^{(4)} = - \hspace{1mm}\parbox{10mm}{\includegraphics[scale=0.09]{fig29.eps}}  -  \frac{1}{2}\hspace{1mm}\parbox{10mm}{\includegraphics[scale=1.0]{fig10.eps}} + 2 \hspace{1mm} perm. - \frac{1}{4}\hspace{1mm}\parbox{16mm}{\includegraphics[scale=1.0]{fig11.eps}} + 2 \hspace{1mm} perm. -  \nonumber \\ && \frac{1}{2}\hspace{1mm}\parbox{12mm}{\includegraphics[scale=0.8]{fig21.eps}} + 5 \hspace{1mm} perm. -  \mathcal{K}
  \left(\parbox{10mm}{\includegraphics[scale=1.0]{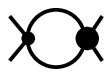}} + 2 \hspace{1mm} perm. \right),
\end{eqnarray}
\begin{eqnarray}\label{gtfrdesuuji}
\Gamma^{(2,1)} =  1 - \frac{1}{2}\hspace{1mm}\parbox{14mm}{\includegraphics[scale=1.0]{fig14.eps}}  - \frac{1}{4}\hspace{1mm}\parbox{12mm}{\includegraphics[scale=1.0]{fig16.eps}}  -  \frac{1}{2}\hspace{1mm}\parbox{12mm}{\includegraphics[scale=0.8]{fig17.eps}} - \frac{1}{2} \mathcal{K}
  \left(\parbox{12mm}{\includegraphics[scale=.2]{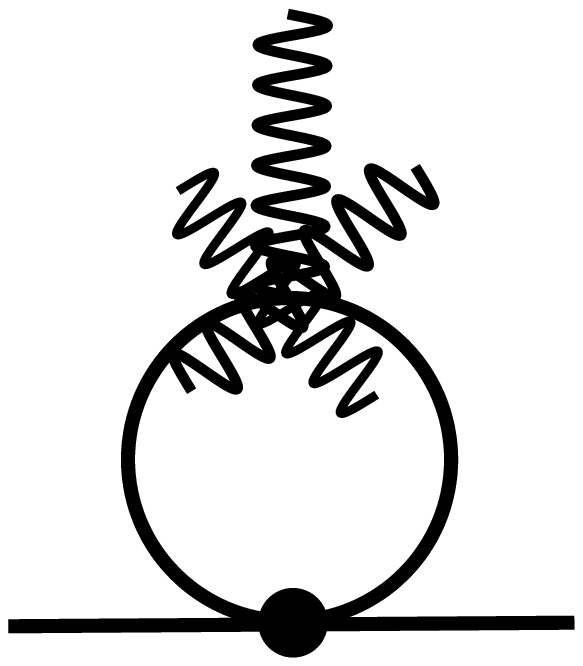}} \right) - \frac{1}{2} \mathcal{K}
  \left(\parbox{12mm}{\includegraphics[scale=.2]{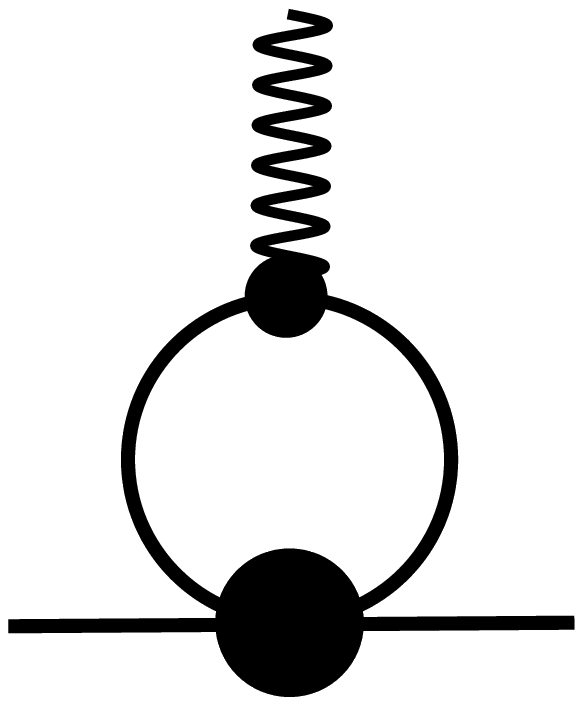}} \right)
\end{eqnarray}
which was attained through the introduction of counterterm diagrams \cite{Kleinert,BogoliubovParasyuk,Hepp,Zimmermann}. As it is known, in the massless theory, we need just a minimal set of Feynman diagrams \cite{Amit}. They are the ones
\begin{eqnarray}
\parbox{10mm}{\includegraphics[scale=1.0]{fig10.eps}} =  \frac{\mu^{\epsilon}}{\epsilon} \left[1 - \frac{1}{2}\epsilon - \frac{1}{2}\epsilon L\Bigg(\frac{P^{2} + K_{\mu\nu}P^{\mu}P^{\nu}}{\mu^{2}}\Bigg) \right]q^{2}\mathbf{\Pi}u^{2},
\end{eqnarray}
\begin{eqnarray}
\parbox{12mm}{\includegraphics[scale=1.0]{fig6.eps}} =  -\frac{(P^{2} + K_{\mu\nu}P^{\mu}P^{\nu})}{8\epsilon}\left[ 1 + \frac{1}{4}\epsilon -2\epsilon L_{3}\Bigg(\frac{P^{2} + K_{\mu\nu}P^{\mu}P^{\nu}}{\mu^{2}}\Bigg) \right] q^{3}\mathbf{\Pi}^{2}u^{2},
\end{eqnarray}  
\begin{eqnarray}
\parbox{10mm}{\includegraphics[scale=0.9]{fig7.eps}} = \frac{(P^{2} + K_{\mu\nu}P^{\mu}P^{\nu})}{6\epsilon^{2}}\left[ 1 + \frac{1}{2}\epsilon -3\epsilon L_{3}\Bigg(\frac{P^{2} + K_{\mu\nu}P^{\mu}P^{\nu}}{\mu^{2}}\Bigg) \right] q^{5}\mathbf{\Pi}^{3}u^{3},
\end{eqnarray}      
\begin{eqnarray}
\parbox{12mm}{\includegraphics[scale=0.8]{fig21.eps}} =  -\frac{\mu^{\epsilon}}{2\epsilon^{2}} \left[1 - \frac{1}{2}\epsilon - \epsilon L\Bigg(\frac{P^{2} + K_{\mu\nu}P^{\mu}P^{\nu}}{\mu^{2}}\Bigg) \right] q^{4}\mathbf{\Pi}^{2}u^{3}.
\end{eqnarray}  
Then we obtain 
\begin{eqnarray}\label{rygyfggfffggyt}
\beta_{q}(u) = -\epsilon u +   \frac{N + 8}{6}q^{2}\mathbf{\Pi}u^{2} -  \frac{3N + 14}{12}q^{4}\mathbf{\Pi}^{2}u^{3} + \frac{N + 2}{36}q^{3}(1-q)\mathbf{\Pi}^{2}u^{3}, 
\end{eqnarray}
\begin{eqnarray}\label{hkfusdfdrs}
\gamma_{\phi ,q} = \frac{N + 2}{72}q^{3}\mathbf{\Pi}^{2}u^{2} - \frac{(N + 2)(N + 8)}{1728}q^{5}\mathbf{\Pi}^{3}u^{3},  
\end{eqnarray}
\begin{eqnarray}\label{srdsdfrdsr}
\gamma_{\phi^{2}, q}(u) = \frac{N + 2}{6}q^{2}\mathbf{\Pi}u -  \frac{5(N + 2)}{72}q^{4}\mathbf{\Pi}^{2}u^{2} + \frac{N + 2}{72}q^{3}(1-q)\mathbf{\Pi}^{2}u^{2},
\end{eqnarray}
where we have to compute $\gamma_{\phi^{2}, q}(u)$ instead $\overline{\gamma}_{\phi^{2}, q}(u)$ in this method and
\begin{eqnarray}
u_{q}^{*} = \frac{6\epsilon}{(N + 8)q^{2}\mathbf{\Pi}}\Bigg\{ 1 +  \epsilon\Bigg[ \frac{3(3N + 14)}{(N + 8)^{2}} -\frac{N + 2}{(N + 8)^{2}}\frac{1 - q}{q} \Bigg]\Bigg\}.
\end{eqnarray}

\subsection{NLO Lorentz-violating $q$-deformed critical exponents} 

\par We are now in a position to compute the NLO radiative quantum corrections to the Lorentz-violating $q$-deformed critical exponents. As there are six such critical indices and four scaling relations among them \cite{Stanley}, we can evaluate two of them independently. By applying the relations $\eta_{q}\equiv\gamma_{\phi,q}(u_{q}^{*})$ and $\nu_{q}^{-1}\equiv 2 - \eta_{q} - \overline{\gamma}_{\phi^{2},q}(u_{q}^{*})$ in the first two methods and $\eta_{q}\equiv\gamma_{\phi,q}(u_{q}^{*})$ and $\nu_{q}^{-1}\equiv 2 - \gamma_{\phi^{2},q}(u_{q}^{*})$ for the last one, we choice to compute independently, $\eta_{q}$ and $\nu_{q}$ and then to evaluate the remaining ones through the scaling relations. Thus we obtain the ones
\begin{eqnarray}
\alpha_{q} = \frac{(4 - N)}{4(N + 8)}\epsilon +  \frac{(N + 2)(N^{2} + 30N + 56)}{4(N + 8)^{3}}\epsilon^{2} - \frac{(N + 2)(4 - N)}{2(N + 8)^{3}}\frac{(1 - q)}{q}\epsilon^{2},
\end{eqnarray}
\begin{eqnarray}
\beta_{q} = \frac{1}{2} - \frac{3}{2(N + 8)}\epsilon +  \frac{(N + 2)(2N + 1)}{2(N + 8)^{3}}\epsilon^{2} + \frac{3(N + 2)}{2(N + 8)^{3}}\frac{(1 - q)}{q}\epsilon^{2},
\end{eqnarray}
\begin{eqnarray}
\gamma_{q} = 1 + \frac{(N + 2)}{2(N + 8)}\epsilon +  \frac{(N + 2)(N^{2} + 22N + 52)}{4(N + 8)^{3}}\epsilon^{2} - \frac{(N + 2)^{2}}{2(N + 8)^{3}}\frac{(1 - q)}{q}\epsilon^{2},
\end{eqnarray}
\begin{eqnarray}
\delta_{q} = 3 +  \epsilon +  \frac{N^{2} + 14N + 60}{2(N + 8)^{2}}\epsilon^{2} - \frac{N + 2}{(N + 8)^{2}}\frac{(1 - q)}{q}\epsilon^{2},
\end{eqnarray}
\begin{eqnarray}
\nu_{q} = \frac{1}{2} + \frac{(N + 2)}{4(N + 8)}\epsilon +  \frac{(N + 2)(N^{2} + 23N + 60)}{8(N + 8)^{3}}\epsilon^{2} + \frac{(N + 2)(4 - N)}{8(N + 8)^{3}}\frac{(1 - q)}{q}\epsilon^{2},
\end{eqnarray}
\begin{eqnarray}
\eta_{q} = \frac{(N + 2)}{2(N + 8)^{2}q}\epsilon^{2}\left\{ 1 + \left[ \frac{6(3N + 14)}{(N + 8)^{2}} -\frac{1}{4} - \frac{2(N + 2)}{(N + 8)^{2}}\frac{(1 - q)}{q} \right]\epsilon\right\},
\end{eqnarray}
which are the same as their corresponding Lorentz-invariant $q$-deformed ones \cite{PhysRevD.97.105006}.

\section{Any loop $q$-deformed critical exponents}

\par In the present section we have to generalize the results of the earlier ones to any loop order. As the Lorentz-violating $q$-deformed critical exponents are universal quantities, we can evaluate their all-loop radiative quantum corrections through any renormalization group scheme. Then by applying a general theorem \cite{PhysRevD.96.116002} through the BPHZ method, we can write the computed expression for some arbitrary Feynman diagram of any loop order. It can be expressed as $\mathbf{\Pi}^{L}\mathcal{F}_{q}(u,P^{2} + K_{\mu\nu}P^{\mu}P^{\nu},\epsilon,\mu)$ if its corresponding Lorentz-invariant counterpart is given by $\mathcal{F}(u,P^{2},\epsilon,\mu)$, where $L$ is the number of loops of the referred diagram. As the momentum-dependence of Lorentz-violating $q$-deformed diagrams is similar to the corresponding Lorentz-invariant ones, \emph{i. e.} $P^{2} + K_{\mu\nu}P^{\mu}P^{\nu}$ corresponds to $P^{2}$, all Lorentz-violating $q$-deformed momentum-dependent integrals cancel out in the middle of calculations as the BPHZ method demands \cite{BogoliubovParasyuk,Hepp,Zimmermann} in the same fashion as in the Lorentz-invariant $q$-deformed case. Thus, we can write the $\beta_{q}$ function and anomalous dimensions for any loop level as
\begin{eqnarray}\label{uhgufhduhufdhu}
\beta_{q}(u) =  -\epsilon u + \sum_{n=2}^{\infty}\beta_{q,n}^{(0)}\mathbf{\Pi}^{n-1}u^{n}, 
\end{eqnarray}
\begin{eqnarray}
\gamma_{q}(u) = \sum_{n=2}^{\infty}\gamma_{q,n}^{(0)}\mathbf{\Pi}^{n}u^{n},
\end{eqnarray}
\begin{eqnarray}
\gamma_{\phi^{2},q}(u) = \sum_{n=1}^{\infty}\gamma_{\phi^{2},q,n}^{(0)}\mathbf{\Pi}^{n}u^{n},
\end{eqnarray}
where the factors $\beta_{q,n}^{(0)}$, $\gamma_{q,n}^{(0)}$ and $\gamma_{\phi^{2},q,n}^{(0)}$ are the corresponding any loop Lorentz-violating $q$-deformed radiative quantum corrections to the referred functions. Now we can factor a power of $u$ from the all-loop order Lorentz-violating $q$-deformed $\beta_{q}$-function and evaluate the all-loop Lorentz-violating $q$-deformed nontrivial fixed point whose solution is given by $u_{q}^{*} = u_{q}^{*(0)}/\mathbf{\Pi}$, where $ u_{q}^{*(0)}$ is the all-loop Lorentz-invariant $q$-deformed nontrivial fixed point. Thus, by computing the Lorentz-violating $q$-deformed anomalous dimensions at this fixed point, we obtain that the all-loop order Lorentz-violating $q$-deformed critical exponents exponents are the same as their all-loop level Lorentz-invariant $q$-deformed counterparts. Once again, we have shown that a symmetry breaking mechanism, the Lorentz-violating one, does not affect the $q$-deformed critical exponents values but now for any loop levels, since this mechanism is one that occurs in the spacetime where the field is embedded and not in its internal one. This fact confirms the universality hypothesis for all loop orders. This completes our generalization procedure. We have now to present our conclusions.

\section{Conclusions}

\par We have probed the effect of the Lorentz-violating symmetry breaking mechanism on the values of the radiative quantum corrections to the critical exponents for massless $q$-deformed O($N$) $\lambda\phi^{4}$ scalar field theories. The violation of the Lorentz symmetry was treated exactly. As the critical exponents are universal quantities, they must present the same results if computed at different renormalization schemes. Thus we have employed three distinct and independent field-theoretic renormalization group methods for attaining that goal. We have initially evaluated the Lorentz-violating $q$-deformed critical exponents up to next-to-leading order. A further step was the generalization of the former task for any loop level. We have shown that the Lorentz-violating $q$-deformed critical exponents turned out to be the same as their Lorentz-invariant $q$-deformed counterparts. This result has confirmed the the universality hypothesis at all-loop order since the symmetry breaking mechanism approached here was not one occurring in the internal space of the field but in the spacetime where the field is embedded. We believe that this symmetry breaking mechanism can be also probed in further works involving the computation of Lorentz-violating $q$-deformed finite size scaling effects, correction to scaling, amplitude ratios etc.

\section*{Acknowledgements}

\par PRSC would like to thank CAPES (brazilian funding agency) for financial support.

\bibliography{apstemplate}

\providecommand{\noopsort}[1]{}\providecommand{\singleletter}[1]{#1}%
\begin{thebibliography}{69}%
\makeatletter
\providecommand \@ifxundefined [1]{%
 \@ifx{#1\undefined}
}%
\providecommand \@ifnum [1]{%
 \ifnum #1\expandafter \@firstoftwo
 \else \expandafter \@secondoftwo
 \fi
}%
\providecommand \@ifx [1]{%
 \ifx #1\expandafter \@firstoftwo
 \else \expandafter \@secondoftwo
 \fi
}%
\providecommand \natexlab [1]{#1}%
\providecommand \enquote  [1]{``#1''}%
\providecommand \bibnamefont  [1]{#1}%
\providecommand \bibfnamefont [1]{#1}%
\providecommand \citenamefont [1]{#1}%
\providecommand \href@noop [0]{\@secondoftwo}%
\providecommand \href [0]{\begingroup \@sanitize@url \@href}%
\providecommand \@href[1]{\@@startlink{#1}\@@href}%
\providecommand \@@href[1]{\endgroup#1\@@endlink}%
\providecommand \@sanitize@url [0]{\catcode `\\12\catcode `\$12\catcode
  `\&12\catcode `\#12\catcode `\^12\catcode `\_12\catcode `\%12\relax}%
\providecommand \@@startlink[1]{}%
\providecommand \@@endlink[0]{}%
\providecommand \url  [0]{\begingroup\@sanitize@url \@url }%
\providecommand \@url [1]{\endgroup\@href {#1}{\urlprefix }}%
\providecommand \urlprefix  [0]{URL }%
\providecommand \Eprint [0]{\href }%
\providecommand \doibase [0]{http://dx.doi.org/}%
\providecommand \selectlanguage [0]{\@gobble}%
\providecommand \bibinfo  [0]{\@secondoftwo}%
\providecommand \bibfield  [0]{\@secondoftwo}%
\providecommand \translation [1]{[#1]}%
\providecommand \BibitemOpen [0]{}%
\providecommand \bibitemStop [0]{}%
\providecommand \bibitemNoStop [0]{.\EOS\space}%
\providecommand \EOS [0]{\spacefactor3000\relax}%
\providecommand \BibitemShut  [1]{\csname bibitem#1\endcsname}%
\let\auto@bib@innerbib\@empty
\bibitem [{\citenamefont {Livine}(2017)}]{Livine2017}%
  \BibitemOpen
  \bibfield  {author} {\bibinfo {author} {\bibfnamefont {E.~R.}\ \bibnamefont
  {Livine}},\ }\href@noop {} {\bibfield  {journal} {\bibinfo  {journal} {Ann.
  Henri Poincar{\'e}}\ }\textbf {\bibinfo {volume} {18}},\ \bibinfo {pages}
  {1465} (\bibinfo {year} {2017})}\BibitemShut {NoStop}%
\bibitem [{\citenamefont {Dil}(2016)}]{Dil}%
  \BibitemOpen
  \bibfield  {author} {\bibinfo {author} {\bibfnamefont {E.}~\bibnamefont
  {Dil}},\ }\href@noop {} {\bibfield  {journal} {\bibinfo  {journal} {Adv. High
  Energy Phys.}\ }\textbf {\bibinfo {volume} {2016}},\ \bibinfo {pages}
  {Article ID 7380372} (\bibinfo {year} {2016})}\BibitemShut {NoStop}%
\bibitem [{\citenamefont {Majid}\ and\ \citenamefont
  {Schroers}(2009)}]{1751-8121-42-42-425402}%
  \BibitemOpen
  \bibfield  {author} {\bibinfo {author} {\bibfnamefont {S.}~\bibnamefont
  {Majid}}\ and\ \bibinfo {author} {\bibfnamefont {B.~J.}\ \bibnamefont
  {Schroers}},\ }\href {http://stacks.iop.org/1751-8121/42/i=42/a=425402}
  {\bibfield  {journal} {\bibinfo  {journal} {J. Phys. A}\ }\textbf {\bibinfo
  {volume} {42}},\ \bibinfo {pages} {425402} (\bibinfo {year}
  {2009})}\BibitemShut {NoStop}%
\bibitem [{\citenamefont {Khavkine}\ and\ \citenamefont
  {Christensen}(2007)}]{0264-9381-24-13-009}%
  \BibitemOpen
  \bibfield  {author} {\bibinfo {author} {\bibfnamefont {I.}~\bibnamefont
  {Khavkine}}\ and\ \bibinfo {author} {\bibfnamefont {J.~D.}\ \bibnamefont
  {Christensen}},\ }\href {http://stacks.iop.org/0264-9381/24/i=13/a=009}
  {\bibfield  {journal} {\bibinfo  {journal} {Class. Quantum Grav.}\ }\textbf
  {\bibinfo {volume} {24}},\ \bibinfo {pages} {3271} (\bibinfo {year}
  {2007})}\BibitemShut {NoStop}%
\bibitem [{\citenamefont {Major}\ and\ \citenamefont
  {Smolin}(1996)}]{MAJOR1996267}%
  \BibitemOpen
  \bibfield  {author} {\bibinfo {author} {\bibfnamefont {S.}~\bibnamefont
  {Major}}\ and\ \bibinfo {author} {\bibfnamefont {L.}~\bibnamefont {Smolin}},\
  }\href {\doibase https://doi.org/10.1016/0550-3213(96)00259-3} {\bibfield
  {journal} {\bibinfo  {journal} {Nucl. Phys. B}\ }\textbf {\bibinfo {volume}
  {473}},\ \bibinfo {pages} {267 } (\bibinfo {year} {1996})}\BibitemShut
  {NoStop}%
\bibitem [{\citenamefont {Gadde}\ \emph {et~al.}(2011)\citenamefont {Gadde},
  \citenamefont {Rastelli}, \citenamefont {Razamat},\ and\ \citenamefont
  {Yan}}]{PhysRevLett.106.241602}%
  \BibitemOpen
  \bibfield  {author} {\bibinfo {author} {\bibfnamefont {A.}~\bibnamefont
  {Gadde}}, \bibinfo {author} {\bibfnamefont {L.}~\bibnamefont {Rastelli}},
  \bibinfo {author} {\bibfnamefont {S.~S.}\ \bibnamefont {Razamat}}, \ and\
  \bibinfo {author} {\bibfnamefont {W.}~\bibnamefont {Yan}},\ }\href {\doibase
  10.1103/PhysRevLett.106.241602} {\bibfield  {journal} {\bibinfo  {journal}
  {Phys. Rev. Lett.}\ }\textbf {\bibinfo {volume} {106}},\ \bibinfo {pages}
  {241602} (\bibinfo {year} {2011})}\BibitemShut {NoStop}%
\bibitem [{\citenamefont {Aganagic}\ \emph {et~al.}(2005)\citenamefont
  {Aganagic}, \citenamefont {Ooguri}, \citenamefont {Saulina},\ and\
  \citenamefont {Vafa}}]{AGANAGIC2005304}%
  \BibitemOpen
  \bibfield  {author} {\bibinfo {author} {\bibfnamefont {M.}~\bibnamefont
  {Aganagic}}, \bibinfo {author} {\bibfnamefont {H.}~\bibnamefont {Ooguri}},
  \bibinfo {author} {\bibfnamefont {N.}~\bibnamefont {Saulina}}, \ and\
  \bibinfo {author} {\bibfnamefont {C.}~\bibnamefont {Vafa}},\ }\href {\doibase
  https://doi.org/10.1016/j.nuclphysb.2005.02.035} {\bibfield  {journal}
  {\bibinfo  {journal} {Nucl. Phys. B}\ }\textbf {\bibinfo {volume} {715}},\
  \bibinfo {pages} {304 } (\bibinfo {year} {2005})}\BibitemShut {NoStop}%
\bibitem [{\citenamefont {Lavagno}\ and\ \citenamefont
  {Swamy}(2000)}]{PhysRevE.61.1218}%
  \BibitemOpen
  \bibfield  {author} {\bibinfo {author} {\bibfnamefont {A.}~\bibnamefont
  {Lavagno}}\ and\ \bibinfo {author} {\bibfnamefont {P.~N.}\ \bibnamefont
  {Swamy}},\ }\href {\doibase 10.1103/PhysRevE.61.1218} {\bibfield  {journal}
  {\bibinfo  {journal} {Phys. Rev. E}\ }\textbf {\bibinfo {volume} {61}},\
  \bibinfo {pages} {1218} (\bibinfo {year} {2000})}\BibitemShut {NoStop}%
\bibitem [{\citenamefont {Delduc}\ \emph {et~al.}(2014)\citenamefont {Delduc},
  \citenamefont {Magro},\ and\ \citenamefont {Vicedo}}]{Delduc2014}%
  \BibitemOpen
  \bibfield  {author} {\bibinfo {author} {\bibfnamefont {F.}~\bibnamefont
  {Delduc}}, \bibinfo {author} {\bibfnamefont {M.}~\bibnamefont {Magro}}, \
  and\ \bibinfo {author} {\bibfnamefont {B.}~\bibnamefont {Vicedo}},\ }\href
  {\doibase 10.1007/JHEP10(2014)132} {\bibfield  {journal} {\bibinfo  {journal}
  {JHEP}\ }\textbf {\bibinfo {volume} {2014}},\ \bibinfo {pages} {132}
  (\bibinfo {year} {2014})}\BibitemShut {NoStop}%
\bibitem [{\citenamefont {Liu}\ \emph {et~al.}(2001)\citenamefont {Liu},
  \citenamefont {Sun}, \citenamefont {Yu},\ and\ \citenamefont
  {Zhou}}]{PhysRevA.63.023802}%
  \BibitemOpen
  \bibfield  {author} {\bibinfo {author} {\bibfnamefont {Y.-X.}\ \bibnamefont
  {Liu}}, \bibinfo {author} {\bibfnamefont {C.~P.}\ \bibnamefont {Sun}},
  \bibinfo {author} {\bibfnamefont {S.~X.}\ \bibnamefont {Yu}}, \ and\ \bibinfo
  {author} {\bibfnamefont {D.~L.}\ \bibnamefont {Zhou}},\ }\href {\doibase
  10.1103/PhysRevA.63.023802} {\bibfield  {journal} {\bibinfo  {journal} {Phys.
  Rev. A}\ }\textbf {\bibinfo {volume} {63}},\ \bibinfo {pages} {023802}
  (\bibinfo {year} {2001})}\BibitemShut {NoStop}%
\bibitem [{\citenamefont {Quesne}(2002)}]{0305-4470-35-43-316}%
  \BibitemOpen
  \bibfield  {author} {\bibinfo {author} {\bibfnamefont {C.}~\bibnamefont
  {Quesne}},\ }\href {http://stacks.iop.org/0305-4470/35/i=43/a=316} {\bibfield
   {journal} {\bibinfo  {journal} {J. Phys. A}\ }\textbf {\bibinfo {volume}
  {35}},\ \bibinfo {pages} {9213} (\bibinfo {year} {2002})}\BibitemShut
  {NoStop}%
\bibitem [{\citenamefont {Conroy}\ \emph {et~al.}(2010)\citenamefont {Conroy},
  \citenamefont {Miller},\ and\ \citenamefont {Plastino}}]{CONROY20104581}%
  \BibitemOpen
  \bibfield  {author} {\bibinfo {author} {\bibfnamefont {J.}~\bibnamefont
  {Conroy}}, \bibinfo {author} {\bibfnamefont {H.}~\bibnamefont {Miller}}, \
  and\ \bibinfo {author} {\bibfnamefont {A.}~\bibnamefont {Plastino}},\ }\href
  {\doibase https://doi.org/10.1016/j.physleta.2010.09.038} {\bibfield
  {journal} {\bibinfo  {journal} {Phys. Lett. A}\ }\textbf {\bibinfo {volume}
  {374}},\ \bibinfo {pages} {4581 } (\bibinfo {year} {2010})}\BibitemShut
  {NoStop}%
\bibitem [{\citenamefont {Griguolo}\ \emph {et~al.}(2007)\citenamefont
  {Griguolo}, \citenamefont {Seminara}, \citenamefont {Szabo},\ and\
  \citenamefont {Tanzini}}]{GRIGUOLO20071}%
  \BibitemOpen
  \bibfield  {author} {\bibinfo {author} {\bibfnamefont {L.}~\bibnamefont
  {Griguolo}}, \bibinfo {author} {\bibfnamefont {D.}~\bibnamefont {Seminara}},
  \bibinfo {author} {\bibfnamefont {R.~J.}\ \bibnamefont {Szabo}}, \ and\
  \bibinfo {author} {\bibfnamefont {A.}~\bibnamefont {Tanzini}},\ }\href
  {\doibase https://doi.org/10.1016/j.nuclphysb.2007.02.030} {\bibfield
  {journal} {\bibinfo  {journal} {Nucl. Phys. B}\ }\textbf {\bibinfo {volume}
  {772}},\ \bibinfo {pages} {1 } (\bibinfo {year} {2007})}\BibitemShut
  {NoStop}%
\bibitem [{\citenamefont {Baseilhac}(2006)}]{BASEILHAC2006309}%
  \BibitemOpen
  \bibfield  {author} {\bibinfo {author} {\bibfnamefont {P.}~\bibnamefont
  {Baseilhac}},\ }\href {\doibase
  https://doi.org/10.1016/j.nuclphysb.2006.08.008} {\bibfield  {journal}
  {\bibinfo  {journal} {Nucl. Phys. B}\ }\textbf {\bibinfo {volume} {754}},\
  \bibinfo {pages} {309 } (\bibinfo {year} {2006})}\BibitemShut {NoStop}%
\bibitem [{\citenamefont {Caporaso}\ \emph {et~al.}(2006)\citenamefont
  {Caporaso}, \citenamefont {Cirafici}, \citenamefont {Griguolo}, \citenamefont
  {Pasquetti}, \citenamefont {Seminara},\ and\ \citenamefont
  {Szabo}}]{1126-6708-2006-01-035}%
  \BibitemOpen
  \bibfield  {author} {\bibinfo {author} {\bibfnamefont {N.}~\bibnamefont
  {Caporaso}}, \bibinfo {author} {\bibfnamefont {M.}~\bibnamefont {Cirafici}},
  \bibinfo {author} {\bibfnamefont {L.}~\bibnamefont {Griguolo}}, \bibinfo
  {author} {\bibfnamefont {S.}~\bibnamefont {Pasquetti}}, \bibinfo {author}
  {\bibfnamefont {D.}~\bibnamefont {Seminara}}, \ and\ \bibinfo {author}
  {\bibfnamefont {R.~J.}\ \bibnamefont {Szabo}},\ }\href
  {http://stacks.iop.org/1126-6708/2006/i=01/a=035} {\bibfield  {journal}
  {\bibinfo  {journal} {JHEP}\ }\textbf {\bibinfo {volume} {2006}},\ \bibinfo
  {pages} {035} (\bibinfo {year} {2006})}\BibitemShut {NoStop}%
\bibitem [{\citenamefont {Sobhani}\ and\ \citenamefont
  {Hassanabadi}(2017)}]{Adv.High.EnergyPhys.20179530874}%
  \BibitemOpen
  \bibfield  {author} {\bibinfo {author} {\bibfnamefont {C.~W.~S.}\
  \bibnamefont {Sobhani}, \bibfnamefont {H.}}\ and\ \bibinfo {author}
  {\bibfnamefont {H.}~\bibnamefont {Hassanabadi}},\ }\href@noop {} {\bibfield
  {journal} {\bibinfo  {journal} {Adv. High Energy Phys.}\ ,\ \bibinfo {pages}
  {9530874}} (\bibinfo {year} {2017})}\BibitemShut {NoStop}%
\bibitem [{\citenamefont {Chung}\ and\ \citenamefont
  {Hassanabadi}(2017{\natexlab{a}})}]{Eur.Phys.J.Plus1322017398}%
  \BibitemOpen
  \bibfield  {author} {\bibinfo {author} {\bibfnamefont {S.~H.}\ \bibnamefont
  {Chung}, \bibfnamefont {W.~S.}}\ and\ \bibinfo {author} {\bibfnamefont
  {H.}~\bibnamefont {Hassanabadi}},\ }\href@noop {} {\bibfield  {journal}
  {\bibinfo  {journal} {Eur. Phys. J. Plus}\ }\textbf {\bibinfo {volume}
  {132}},\ \bibinfo {pages} {398} (\bibinfo {year}
  {2017}{\natexlab{a}})}\BibitemShut {NoStop}%
\bibitem [{\citenamefont {Chung}\ and\ \citenamefont
  {Hassanabadi}(2017{\natexlab{b}})}]{Int.J.Theor.Phys.5620171746}%
  \BibitemOpen
  \bibfield  {author} {\bibinfo {author} {\bibfnamefont {W.~S.}\ \bibnamefont
  {Chung}}\ and\ \bibinfo {author} {\bibfnamefont {H.}~\bibnamefont
  {Hassanabadi}},\ }\href@noop {} {\bibfield  {journal} {\bibinfo  {journal}
  {Int. J. Theor. Phys.}\ }\textbf {\bibinfo {volume} {56}},\ \bibinfo {pages}
  {1746} (\bibinfo {year} {2017}{\natexlab{b}})}\BibitemShut {NoStop}%
\bibitem [{\citenamefont {Dil}(2017)}]{Phys.DarkUniv.1620171}%
  \BibitemOpen
  \bibfield  {author} {\bibinfo {author} {\bibfnamefont {E.}~\bibnamefont
  {Dil}},\ }\href@noop {} {\bibfield  {journal} {\bibinfo  {journal} {Phys.
  Dark Univ.}\ }\textbf {\bibinfo {volume} {16}},\ \bibinfo {pages} {1}
  (\bibinfo {year} {2017})}\BibitemShut {NoStop}%
\bibitem [{\citenamefont {Boumali}\ and\ \citenamefont
  {Hassanabadi}(2017)}]{Adv.HighEnergyPhys.20179371391}%
  \BibitemOpen
  \bibfield  {author} {\bibinfo {author} {\bibfnamefont {A.}~\bibnamefont
  {Boumali}}\ and\ \bibinfo {author} {\bibfnamefont {H.}~\bibnamefont
  {Hassanabadi}},\ }\href@noop {} {\bibfield  {journal} {\bibinfo  {journal}
  {Adv. High Energy Phys.}\ ,\ \bibinfo {pages} {9371391}} (\bibinfo {year}
  {2017})}\BibitemShut {NoStop}%
\bibitem [{\citenamefont {Du}\ and\ \citenamefont
  {Zhou}(2016)}]{EPL11320162000}%
  \BibitemOpen
  \bibfield  {author} {\bibinfo {author} {\bibfnamefont {X.~K.}\ \bibnamefont
  {Du}, \bibfnamefont {G.~J.}}\ and\ \bibinfo {author} {\bibfnamefont
  {C.}~\bibnamefont {Zhou}},\ }\href@noop {} {\bibfield  {journal} {\bibinfo
  {journal} {EPL}\ }\textbf {\bibinfo {volume} {113}},\ \bibinfo {pages}
  {20002} (\bibinfo {year} {2016})}\BibitemShut {NoStop}%
\bibitem [{\citenamefont {Watanabe}(2016)}]{JHEP1220160630}%
  \BibitemOpen
  \bibfield  {author} {\bibinfo {author} {\bibfnamefont {N.}~\bibnamefont
  {Watanabe}},\ }\href@noop {} {\bibfield  {journal} {\bibinfo  {journal}
  {JHEP}\ }\textbf {\bibinfo {volume} {12}},\ \bibinfo {pages} {063} (\bibinfo
  {year} {2016})}\BibitemShut {NoStop}%
\bibitem [{\citenamefont {Dey}(2015)}]{Phys.Rev.D912015044024}%
  \BibitemOpen
  \bibfield  {author} {\bibinfo {author} {\bibfnamefont {S.}~\bibnamefont
  {Dey}},\ }\href@noop {} {\bibfield  {journal} {\bibinfo  {journal} {Phys.
  Rev. D}\ }\textbf {\bibinfo {volume} {91}},\ \bibinfo {pages} {044024}
  (\bibinfo {year} {2015})}\BibitemShut {NoStop}%
\bibitem [{\citenamefont {Vinod}(1997)}]{G.Vinod}%
  \BibitemOpen
  \bibfield  {author} {\bibinfo {author} {\bibfnamefont {G.}~\bibnamefont
  {Vinod}},\ }\href@noop {} {\emph {\bibinfo {title} {PhD Thesis}}}\ (\bibinfo
  {publisher} {Cochin University of Science and Technology, Department of
  Physics},\ \bibinfo {year} {1997})\BibitemShut {NoStop}%
\bibitem [{\citenamefont {Carvalho}(2018)}]{PhysRevD.97.105006}%
  \BibitemOpen
  \bibfield  {author} {\bibinfo {author} {\bibfnamefont {P.~R.~S.}\
  \bibnamefont {Carvalho}},\ }\href {\doibase 10.1103/PhysRevD.97.105006}
  {\bibfield  {journal} {\bibinfo  {journal} {Phys. Rev. D}\ }\textbf {\bibinfo
  {volume} {97}},\ \bibinfo {pages} {105006} (\bibinfo {year}
  {2018})}\BibitemShut {NoStop}%
\bibitem [{\citenamefont {Stanley}(1988)}]{Stanley}%
  \BibitemOpen
  \bibfield  {author} {\bibinfo {author} {\bibfnamefont {H.~E.}\ \bibnamefont
  {Stanley}},\ }\href@noop {} {\emph {\bibinfo {title} {Introduction to Phase
  Transitions and Critical Phenomena}}}\ (\bibinfo  {publisher} {Oxford
  University Pres},\ \bibinfo {year} {1988})\BibitemShut {NoStop}%
\bibitem [{\citenamefont {Edwards}\ and\ \citenamefont
  {Kosteleck\'y}(2018)}]{EDWARDS2018319}%
  \BibitemOpen
  \bibfield  {author} {\bibinfo {author} {\bibfnamefont {B.~R.}\ \bibnamefont
  {Edwards}}\ and\ \bibinfo {author} {\bibfnamefont {V.~A.}\ \bibnamefont
  {Kosteleck\'y}},\ }\href {\doibase
  https://doi.org/10.1016/j.physletb.2018.10.011} {\bibfield  {journal}
  {\bibinfo  {journal} {Phys. Lett. B}\ }\textbf {\bibinfo {volume} {786}},\
  \bibinfo {pages} {319 } (\bibinfo {year} {2018})}\BibitemShut {NoStop}%
\bibitem [{\citenamefont {Borges}\ \emph {et~al.}()\citenamefont {Borges},
  \citenamefont {Ferrari},\ and\ \citenamefont
  {Barone}}]{Borges.Ferrari.Barone}%
  \BibitemOpen
  \bibfield  {author} {\bibinfo {author} {\bibfnamefont {L.}~\bibnamefont
  {Borges}}, \bibinfo {author} {\bibfnamefont {A.}~\bibnamefont {Ferrari}}, \
  and\ \bibinfo {author} {\bibfnamefont {F.}~\bibnamefont {Barone}},\
  }\href@noop {} {\bibinfo  {journal} {arXiv:1809.08883}\ }\BibitemShut
  {NoStop}%
\bibitem [{\citenamefont {Xiao}(2018)}]{PhysRevD.98.035018}%
  \BibitemOpen
\bibfield  {journal} {  }\bibfield  {author} {\bibinfo {author} {\bibfnamefont
  {Z.}~\bibnamefont {Xiao}},\ }\href {\doibase 10.1103/PhysRevD.98.035018}
  {\bibfield  {journal} {\bibinfo  {journal} {Phys. Rev. D}\ }\textbf {\bibinfo
  {volume} {98}},\ \bibinfo {pages} {035018} (\bibinfo {year}
  {2018})}\BibitemShut {NoStop}%
\bibitem [{\citenamefont {de~Paula~Netto}(2018)}]{PhysRevD.97.055048}%
  \BibitemOpen
  \bibfield  {author} {\bibinfo {author} {\bibfnamefont {T.}~\bibnamefont
  {de~Paula~Netto}},\ }\href {\doibase 10.1103/PhysRevD.97.055048} {\bibfield
  {journal} {\bibinfo  {journal} {Phys. Rev. D}\ }\textbf {\bibinfo {volume}
  {97}},\ \bibinfo {pages} {055048} (\bibinfo {year} {2018})}\BibitemShut
  {NoStop}%
\bibitem [{\citenamefont {Vieira}\ and\ \citenamefont
  {de~Carvalho}(2016)}]{Int.J.Geom.MethodsMod.Phys.13.1650049}%
  \BibitemOpen
  \bibfield  {author} {\bibinfo {author} {\bibfnamefont {W.~d.~C.}\
  \bibnamefont {Vieira}}\ and\ \bibinfo {author} {\bibfnamefont {P.~R.~S.}\
  \bibnamefont {de~Carvalho}},\ }\href@noop {} {\bibfield  {journal} {\bibinfo
  {journal} {Int. J. Geom. Methods Mod. Phys.}\ }\textbf {\bibinfo {volume}
  {13}},\ \bibinfo {pages} {1650049} (\bibinfo {year} {2016})}\BibitemShut
  {NoStop}%
\bibitem [{\citenamefont {Silva}\ and\ \citenamefont
  {Carvalho}(2018)}]{doi:10.1142/S021988781850086X}%
  \BibitemOpen
  \bibfield  {author} {\bibinfo {author} {\bibfnamefont {G.~S.}\ \bibnamefont
  {Silva}}\ and\ \bibinfo {author} {\bibfnamefont {P.~R.~S.}\ \bibnamefont
  {Carvalho}},\ }\href@noop {} {\bibfield  {journal} {\bibinfo  {journal} {Int.
  J. Geom. Methods Mod. Phys.}\ }\textbf {\bibinfo {volume} {15}},\ \bibinfo
  {pages} {1850086} (\bibinfo {year} {2018})}\BibitemShut {NoStop}%
\bibitem [{\citenamefont {Baeta~Scarpelli}\ \emph {et~al.}(2017)\citenamefont
  {Baeta~Scarpelli}, \citenamefont {Brito}, \citenamefont {Felipe},
  \citenamefont {Nascimento},\ and\ \citenamefont
  {Petrov}}]{BaetaScarpelli2017}%
  \BibitemOpen
  \bibfield  {author} {\bibinfo {author} {\bibfnamefont {A.~P.}\ \bibnamefont
  {Baeta~Scarpelli}}, \bibinfo {author} {\bibfnamefont {L.~C.~T.}\ \bibnamefont
  {Brito}}, \bibinfo {author} {\bibfnamefont {J.~C.~C.}\ \bibnamefont
  {Felipe}}, \bibinfo {author} {\bibfnamefont {J.~R.}\ \bibnamefont
  {Nascimento}}, \ and\ \bibinfo {author} {\bibfnamefont {A.~Y.}\ \bibnamefont
  {Petrov}},\ }\href {\doibase 10.1140/epjc/s10052-017-5430-4} {\bibfield
  {journal} {\bibinfo  {journal} {Eur. Phys. J. C}\ }\textbf {\bibinfo {volume}
  {77}},\ \bibinfo {pages} {850} (\bibinfo {year} {2017})}\BibitemShut
  {NoStop}%
\bibitem [{\citenamefont {Cruz}\ \emph {et~al.}(2017)\citenamefont {Cruz},
  \citenamefont {Bezerra~de Mello},\ and\ \citenamefont
  {Petrov}}]{PhysRevD.96.045019}%
  \BibitemOpen
  \bibfield  {author} {\bibinfo {author} {\bibfnamefont {M.~B.}\ \bibnamefont
  {Cruz}}, \bibinfo {author} {\bibfnamefont {E.~R.}\ \bibnamefont {Bezerra~de
  Mello}}, \ and\ \bibinfo {author} {\bibfnamefont {A.~Y.}\ \bibnamefont
  {Petrov}},\ }\href {\doibase 10.1103/PhysRevD.96.045019} {\bibfield
  {journal} {\bibinfo  {journal} {Phys. Rev. D}\ }\textbf {\bibinfo {volume}
  {96}},\ \bibinfo {pages} {045019} (\bibinfo {year} {2017})}\BibitemShut
  {NoStop}%
\bibitem [{\citenamefont {Cruz}\ \emph {et~al.}(2018)\citenamefont {Cruz},
  \citenamefont {Bezerra~de Mello},\ and\ \citenamefont
  {Petrov}}]{doi:10.1142/S0217732318501158}%
  \BibitemOpen
  \bibfield  {author} {\bibinfo {author} {\bibfnamefont {M.~B.}\ \bibnamefont
  {Cruz}}, \bibinfo {author} {\bibfnamefont {E.~R.}\ \bibnamefont {Bezerra~de
  Mello}}, \ and\ \bibinfo {author} {\bibfnamefont {A.~Y.}\ \bibnamefont
  {Petrov}},\ }\href@noop {} {\bibfield  {journal} {\bibinfo  {journal} {Mod.
  Phys. Lett. A}\ }\textbf {\bibinfo {volume} {33}},\ \bibinfo {pages}
  {1850115} (\bibinfo {year} {2018})}\BibitemShut {NoStop}%
\bibitem [{\citenamefont {Kamand}\ \emph {et~al.}(2017)\citenamefont {Kamand},
  \citenamefont {Altschul},\ and\ \citenamefont
  {Schindler}}]{PhysRevD.95.056005}%
  \BibitemOpen
  \bibfield  {author} {\bibinfo {author} {\bibfnamefont {R.}~\bibnamefont
  {Kamand}}, \bibinfo {author} {\bibfnamefont {B.}~\bibnamefont {Altschul}}, \
  and\ \bibinfo {author} {\bibfnamefont {M.~R.}\ \bibnamefont {Schindler}},\
  }\href {\doibase 10.1103/PhysRevD.95.056005} {\bibfield  {journal} {\bibinfo
  {journal} {Phys. Rev. D}\ }\textbf {\bibinfo {volume} {95}},\ \bibinfo
  {pages} {056005} (\bibinfo {year} {2017})}\BibitemShut {NoStop}%
\bibitem [{\citenamefont {Casana}\ and\ \citenamefont
  {da~Silva}(2015)}]{doi:10.1142/S0217732315500376}%
  \BibitemOpen
  \bibfield  {author} {\bibinfo {author} {\bibfnamefont {R.}~\bibnamefont
  {Casana}}\ and\ \bibinfo {author} {\bibfnamefont {K.~A.~T.}\ \bibnamefont
  {da~Silva}},\ }\href@noop {} {\bibfield  {journal} {\bibinfo  {journal} {Mod.
  Phys. Lett. A}\ }\textbf {\bibinfo {volume} {30}},\ \bibinfo {pages}
  {1550037} (\bibinfo {year} {2015})}\BibitemShut {NoStop}%
\bibitem [{\citenamefont {Carvalho}(2013)}]{Carvalho2013850}%
  \BibitemOpen
  \bibfield  {author} {\bibinfo {author} {\bibfnamefont {P.~R.~S.}\
  \bibnamefont {Carvalho}},\ }\href@noop {} {\bibfield  {journal} {\bibinfo
  {journal} {Phys. Lett. B}\ }\textbf {\bibinfo {volume} {726}},\ \bibinfo
  {pages} {850} (\bibinfo {year} {2013})}\BibitemShut {NoStop}%
\bibitem [{\citenamefont {Carvalho}(2014)}]{Carvalho2014320}%
  \BibitemOpen
  \bibfield  {author} {\bibinfo {author} {\bibfnamefont {P.~R.~S.}\
  \bibnamefont {Carvalho}},\ }\href@noop {} {\bibfield  {journal} {\bibinfo
  {journal} {Phys. Lett. B}\ }\textbf {\bibinfo {volume} {730}},\ \bibinfo
  {pages} {320} (\bibinfo {year} {2014})}\BibitemShut {NoStop}%
\bibitem [{\citenamefont {Altschul}(2013)}]{PhysRevD.87.045012}%
  \BibitemOpen
  \bibfield  {author} {\bibinfo {author} {\bibfnamefont {B.}~\bibnamefont
  {Altschul}},\ }\href {\doibase 10.1103/PhysRevD.87.045012} {\bibfield
  {journal} {\bibinfo  {journal} {Phys. Rev. D}\ }\textbf {\bibinfo {volume}
  {87}},\ \bibinfo {pages} {045012} (\bibinfo {year} {2013})}\BibitemShut
  {NoStop}%
\bibitem [{\citenamefont {Ferrero}\ and\ \citenamefont
  {Altschul}(2011)}]{PhysRevD.84.065030}%
  \BibitemOpen
  \bibfield  {author} {\bibinfo {author} {\bibfnamefont {A.}~\bibnamefont
  {Ferrero}}\ and\ \bibinfo {author} {\bibfnamefont {B.}~\bibnamefont
  {Altschul}},\ }\href@noop {} {\bibfield  {journal} {\bibinfo  {journal}
  {Phys. Rev. D}\ }\textbf {\bibinfo {volume} {84}},\ \bibinfo {pages} {065030}
  (\bibinfo {year} {2011})}\BibitemShut {NoStop}%
\bibitem [{\citenamefont {Barreto}\ \emph {et~al.}(2006)\citenamefont
  {Barreto}, \citenamefont {Bazeia},\ and\ \citenamefont
  {Menezes}}]{PhysRevD.73.065015}%
  \BibitemOpen
  \bibfield  {author} {\bibinfo {author} {\bibfnamefont {M.~N.}\ \bibnamefont
  {Barreto}}, \bibinfo {author} {\bibfnamefont {D.}~\bibnamefont {Bazeia}}, \
  and\ \bibinfo {author} {\bibfnamefont {R.}~\bibnamefont {Menezes}},\ }\href
  {\doibase 10.1103/PhysRevD.73.065015} {\bibfield  {journal} {\bibinfo
  {journal} {Phys. Rev. D}\ }\textbf {\bibinfo {volume} {73}},\ \bibinfo
  {pages} {065015} (\bibinfo {year} {2006})}\BibitemShut {NoStop}%
\bibitem [{\citenamefont {Altschul}(2006)}]{ALTSCHUL2006679}%
  \BibitemOpen
  \bibfield  {author} {\bibinfo {author} {\bibfnamefont {B.}~\bibnamefont
  {Altschul}},\ }\href {\doibase
  https://doi.org/10.1016/j.physletb.2006.07.021} {\bibfield  {journal}
  {\bibinfo  {journal} {Phys. Lett. B}\ }\textbf {\bibinfo {volume} {639}},\
  \bibinfo {pages} {679 } (\bibinfo {year} {2006})}\BibitemShut {NoStop}%
\bibitem [{\citenamefont {Anderson}\ \emph {et~al.}(2004)\citenamefont
  {Anderson}, \citenamefont {Sher},\ and\ \citenamefont
  {Turan}}]{PhysRevD.70.016001}%
  \BibitemOpen
  \bibfield  {author} {\bibinfo {author} {\bibfnamefont {D.~L.}\ \bibnamefont
  {Anderson}}, \bibinfo {author} {\bibfnamefont {M.}~\bibnamefont {Sher}}, \
  and\ \bibinfo {author} {\bibfnamefont {I.}~\bibnamefont {Turan}},\ }\href
  {\doibase 10.1103/PhysRevD.70.016001} {\bibfield  {journal} {\bibinfo
  {journal} {Phys. Rev. D}\ }\textbf {\bibinfo {volume} {70}},\ \bibinfo
  {pages} {016001} (\bibinfo {year} {2004})}\BibitemShut {NoStop}%
\bibitem [{\citenamefont {Berger}\ and\ \citenamefont
  {Kosteleck\'y}(2002)}]{PhysRevD.65.091701}%
  \BibitemOpen
  \bibfield  {author} {\bibinfo {author} {\bibfnamefont {M.~S.}\ \bibnamefont
  {Berger}}\ and\ \bibinfo {author} {\bibfnamefont {V.~A.}\ \bibnamefont
  {Kosteleck\'y}},\ }\href {\doibase 10.1103/PhysRevD.65.091701} {\bibfield
  {journal} {\bibinfo  {journal} {Phys. Rev. D}\ }\textbf {\bibinfo {volume}
  {65}},\ \bibinfo {pages} {091701} (\bibinfo {year} {2002})}\BibitemShut
  {NoStop}%
\bibitem [{\citenamefont {Hees}\ \emph {et~al.}(2016)\citenamefont {Hees},
  \citenamefont {Bailey}, \citenamefont {Bourgoin}, \citenamefont
  {Pihan-Le~Bars}, \citenamefont {Guerlin},\ and\ \citenamefont
  {Le~Poncin-Lafitte}}]{universe2201630}%
  \BibitemOpen
  \bibfield  {author} {\bibinfo {author} {\bibfnamefont {A.}~\bibnamefont
  {Hees}}, \bibinfo {author} {\bibfnamefont {Q.~G.}\ \bibnamefont {Bailey}},
  \bibinfo {author} {\bibfnamefont {A.}~\bibnamefont {Bourgoin}}, \bibinfo
  {author} {\bibfnamefont {H.}~\bibnamefont {Pihan-Le~Bars}}, \bibinfo {author}
  {\bibfnamefont {C.}~\bibnamefont {Guerlin}}, \ and\ \bibinfo {author}
  {\bibfnamefont {C.}~\bibnamefont {Le~Poncin-Lafitte}},\ }\href@noop {}
  {\bibfield  {journal} {\bibinfo  {journal} {Universe}\ }\textbf {\bibinfo
  {volume} {2}} (\bibinfo {year} {2016})}\BibitemShut {NoStop}%
\bibitem [{\citenamefont {Kosteleck\'y}\ and\ \citenamefont
  {Samuel}(1989{\natexlab{a}})}]{PhysRevD.40.1886}%
  \BibitemOpen
  \bibfield  {author} {\bibinfo {author} {\bibfnamefont {V.~A.}\ \bibnamefont
  {Kosteleck\'y}}\ and\ \bibinfo {author} {\bibfnamefont {S.}~\bibnamefont
  {Samuel}},\ }\href {\doibase 10.1103/PhysRevD.40.1886} {\bibfield  {journal}
  {\bibinfo  {journal} {Phys. Rev. D}\ }\textbf {\bibinfo {volume} {40}},\
  \bibinfo {pages} {1886} (\bibinfo {year} {1989}{\natexlab{a}})}\BibitemShut
  {NoStop}%
\bibitem [{\citenamefont {Kosteleck\'y}\ and\ \citenamefont
  {Samuel}(1989{\natexlab{b}})}]{PhysRevD.39.683}%
  \BibitemOpen
  \bibfield  {author} {\bibinfo {author} {\bibfnamefont {V.~A.}\ \bibnamefont
  {Kosteleck\'y}}\ and\ \bibinfo {author} {\bibfnamefont {S.}~\bibnamefont
  {Samuel}},\ }\href {\doibase 10.1103/PhysRevD.39.683} {\bibfield  {journal}
  {\bibinfo  {journal} {Phys. Rev. D}\ }\textbf {\bibinfo {volume} {39}},\
  \bibinfo {pages} {683} (\bibinfo {year} {1989}{\natexlab{b}})}\BibitemShut
  {NoStop}%
\bibitem [{\citenamefont {Kosteleck\'y}\ and\ \citenamefont
  {Potting}(1991)}]{Kostelecký1991545}%
  \BibitemOpen
  \bibfield  {author} {\bibinfo {author} {\bibfnamefont {V.}~\bibnamefont
  {Kosteleck\'y}}\ and\ \bibinfo {author} {\bibfnamefont {R.}~\bibnamefont
  {Potting}},\ }\href@noop {} {\bibfield  {journal} {\bibinfo  {journal} {Nucl.
  Phys.}\ }\textbf {\bibinfo {volume} {B359}},\ \bibinfo {pages} {545}
  (\bibinfo {year} {1991})}\BibitemShut {NoStop}%
\bibitem [{\citenamefont {Gambini}\ and\ \citenamefont
  {Pullin}(1999)}]{PhysRevD.59.124021}%
  \BibitemOpen
  \bibfield  {author} {\bibinfo {author} {\bibfnamefont {R.}~\bibnamefont
  {Gambini}}\ and\ \bibinfo {author} {\bibfnamefont {J.}~\bibnamefont
  {Pullin}},\ }\href {\doibase 10.1103/PhysRevD.59.124021} {\bibfield
  {journal} {\bibinfo  {journal} {Phys. Rev. D}\ }\textbf {\bibinfo {volume}
  {59}},\ \bibinfo {pages} {124021} (\bibinfo {year} {1999})}\BibitemShut
  {NoStop}%
\bibitem [{\citenamefont {Amelino-Camelia}(2013)}]{Amelino-Camelia20135}%
  \BibitemOpen
  \bibfield  {author} {\bibinfo {author} {\bibfnamefont {G.}~\bibnamefont
  {Amelino-Camelia}},\ }\href@noop {} {\bibfield  {journal} {\bibinfo
  {journal} {Rel.}\ }\textbf {\bibinfo {volume} {16}},\ \bibinfo {pages} {5}
  (\bibinfo {year} {2013})}\BibitemShut {NoStop}%
\bibitem [{\citenamefont {Myers}\ and\ \citenamefont
  {Pospelov}(2003)}]{PhysRevLett.90.211601}%
  \BibitemOpen
  \bibfield  {author} {\bibinfo {author} {\bibfnamefont {R.~C.}\ \bibnamefont
  {Myers}}\ and\ \bibinfo {author} {\bibfnamefont {M.}~\bibnamefont
  {Pospelov}},\ }\href {\doibase 10.1103/PhysRevLett.90.211601} {\bibfield
  {journal} {\bibinfo  {journal} {Phys. Rev. Lett.}\ }\textbf {\bibinfo
  {volume} {90}},\ \bibinfo {pages} {211601} (\bibinfo {year}
  {2003})}\BibitemShut {NoStop}%
\bibitem [{\citenamefont {Hayakawa}(2000)}]{Hayakawa200039}%
  \BibitemOpen
  \bibfield  {author} {\bibinfo {author} {\bibfnamefont {M.}~\bibnamefont
  {Hayakawa}},\ }\href@noop {} {\bibfield  {journal} {\bibinfo  {journal}
  {Phys. Lett. B}\ }\textbf {\bibinfo {volume} {478}},\ \bibinfo {pages} {39}
  (\bibinfo {year} {2000})}\BibitemShut {NoStop}%
\bibitem [{\citenamefont {Carroll}\ \emph {et~al.}(2001)\citenamefont
  {Carroll}, \citenamefont {Harvey}, \citenamefont {Kosteleck\'y},
  \citenamefont {Lane},\ and\ \citenamefont {Okamoto}}]{PhysRevLett.87.141601}%
  \BibitemOpen
  \bibfield  {author} {\bibinfo {author} {\bibfnamefont {S.~M.}\ \bibnamefont
  {Carroll}}, \bibinfo {author} {\bibfnamefont {J.~A.}\ \bibnamefont {Harvey}},
  \bibinfo {author} {\bibfnamefont {V.~A.}\ \bibnamefont {Kosteleck\'y}},
  \bibinfo {author} {\bibfnamefont {C.~D.}\ \bibnamefont {Lane}}, \ and\
  \bibinfo {author} {\bibfnamefont {T.}~\bibnamefont {Okamoto}},\ }\href
  {\doibase 10.1103/PhysRevLett.87.141601} {\bibfield  {journal} {\bibinfo
  {journal} {Phys. Rev. Lett.}\ }\textbf {\bibinfo {volume} {87}},\ \bibinfo
  {pages} {141601} (\bibinfo {year} {2001})}\BibitemShut {NoStop}%
\bibitem [{\citenamefont {Kislat}\ and\ \citenamefont
  {Krawczynski}(2015)}]{PhysRevD.92.045016}%
  \BibitemOpen
  \bibfield  {author} {\bibinfo {author} {\bibfnamefont {F.}~\bibnamefont
  {Kislat}}\ and\ \bibinfo {author} {\bibfnamefont {H.}~\bibnamefont
  {Krawczynski}},\ }\href {\doibase 10.1103/PhysRevD.92.045016} {\bibfield
  {journal} {\bibinfo  {journal} {Phys. Rev. D}\ }\textbf {\bibinfo {volume}
  {92}},\ \bibinfo {pages} {045016} (\bibinfo {year} {2015})}\BibitemShut
  {NoStop}%
\bibitem [{\citenamefont {Ribeiro}\ \emph {et~al.}(2015)\citenamefont
  {Ribeiro}, \citenamefont {Passos}, \citenamefont {Furtado},\ and\
  \citenamefont {Nascimento}}]{doi:10.1142/S0217751X15500724}%
  \BibitemOpen
  \bibfield  {author} {\bibinfo {author} {\bibfnamefont {L.~R.}\ \bibnamefont
  {Ribeiro}}, \bibinfo {author} {\bibfnamefont {E.}~\bibnamefont {Passos}},
  \bibinfo {author} {\bibfnamefont {C.}~\bibnamefont {Furtado}}, \ and\
  \bibinfo {author} {\bibfnamefont {J.~R.}\ \bibnamefont {Nascimento}},\ }\href
  {\doibase 10.1142/S0217751X15500724} {\bibfield  {journal} {\bibinfo
  {journal} {Int. J. Mod. Phys. A}\ }\textbf {\bibinfo {volume} {30}},\
  \bibinfo {pages} {1550072} (\bibinfo {year} {2015})}\BibitemShut {NoStop}%
\bibitem [{\citenamefont {van Tilburg}\ and\ \citenamefont {van
  Veghel}(2015)}]{VANTILBURG2015236}%
  \BibitemOpen
  \bibfield  {author} {\bibinfo {author} {\bibfnamefont {J.}~\bibnamefont {van
  Tilburg}}\ and\ \bibinfo {author} {\bibfnamefont {M.}~\bibnamefont {van
  Veghel}},\ }\href {\doibase https://doi.org/10.1016/j.physletb.2015.01.036}
  {\bibfield  {journal} {\bibinfo  {journal} {Phys. Lett. B}\ }\textbf
  {\bibinfo {volume} {742}},\ \bibinfo {pages} {236 } (\bibinfo {year}
  {2015})}\BibitemShut {NoStop}%
\bibitem [{\citenamefont {Anacleto}\ \emph {et~al.}(2012)\citenamefont
  {Anacleto}, \citenamefont {Brito},\ and\ \citenamefont
  {Passos}}]{PhysRevD.86.125015}%
  \BibitemOpen
  \bibfield  {author} {\bibinfo {author} {\bibfnamefont {M.~A.}\ \bibnamefont
  {Anacleto}}, \bibinfo {author} {\bibfnamefont {F.~A.}\ \bibnamefont {Brito}},
  \ and\ \bibinfo {author} {\bibfnamefont {E.}~\bibnamefont {Passos}},\ }\href
  {\doibase 10.1103/PhysRevD.86.125015} {\bibfield  {journal} {\bibinfo
  {journal} {Phys. Rev. D}\ }\textbf {\bibinfo {volume} {86}},\ \bibinfo
  {pages} {125015} (\bibinfo {year} {2012})}\BibitemShut {NoStop}%
\bibitem [{\citenamefont {Colladay}\ and\ \citenamefont
  {Kosteleck\'y}(1998)}]{PhysRevD.58.116002}%
  \BibitemOpen
  \bibfield  {author} {\bibinfo {author} {\bibfnamefont {D.}~\bibnamefont
  {Colladay}}\ and\ \bibinfo {author} {\bibfnamefont {V.~A.}\ \bibnamefont
  {Kosteleck\'y}},\ }\href {\doibase 10.1103/PhysRevD.58.116002} {\bibfield
  {journal} {\bibinfo  {journal} {Phys. Rev. D}\ }\textbf {\bibinfo {volume}
  {58}},\ \bibinfo {pages} {116002} (\bibinfo {year} {1998})}\BibitemShut
  {NoStop}%
\bibitem [{\citenamefont {Carvalho}\ and\ \citenamefont
  {Sena-Junior}(2017{\natexlab{a}})}]{CARVALHO2017290}%
  \BibitemOpen
  \bibfield  {author} {\bibinfo {author} {\bibfnamefont {P.~R.~S.}\
  \bibnamefont {Carvalho}}\ and\ \bibinfo {author} {\bibfnamefont {M.~I.}\
  \bibnamefont {Sena-Junior}},\ }\href {\doibase
  https://doi.org/10.1016/j.aop.2017.10.016} {\bibfield  {journal} {\bibinfo
  {journal} {Ann. Phys.}\ }\textbf {\bibinfo {volume} {387}},\ \bibinfo {pages}
  {290} (\bibinfo {year} {2017}{\natexlab{a}})}\BibitemShut {NoStop}%
\bibitem [{\citenamefont {Carvalho}\ and\ \citenamefont
  {Sena-Junior}(2017{\natexlab{b}})}]{Carvalho2017}%
  \BibitemOpen
  \bibfield  {author} {\bibinfo {author} {\bibfnamefont {P.~R.~S.}\
  \bibnamefont {Carvalho}}\ and\ \bibinfo {author} {\bibfnamefont {M.~I.}\
  \bibnamefont {Sena-Junior}},\ }\href {\doibase
  10.1140/epjc/s10052-017-5304-9} {\bibfield  {journal} {\bibinfo  {journal}
  {Eur. Phys. J. C}\ }\textbf {\bibinfo {volume} {77}},\ \bibinfo {pages} {753}
  (\bibinfo {year} {2017}{\natexlab{b}})}\BibitemShut {NoStop}%
\bibitem [{\citenamefont {Wilson}\ and\ \citenamefont
  {Kogut}(1974)}]{Wilson197475}%
  \BibitemOpen
  \bibfield  {author} {\bibinfo {author} {\bibfnamefont {K.~G.}\ \bibnamefont
  {Wilson}}\ and\ \bibinfo {author} {\bibfnamefont {J.}~\bibnamefont {Kogut}},\
  }\href@noop {} {\bibfield  {journal} {\bibinfo  {journal} {Phys. Rep.}\
  }\textbf {\bibinfo {volume} {12}},\ \bibinfo {pages} {75} (\bibinfo {year}
  {1974})}\BibitemShut {NoStop}%
\bibitem [{\citenamefont {Amit}\ and\ \citenamefont
  {Mart\'in-Mayor}(2005)}]{Amit}%
  \BibitemOpen
  \bibfield  {author} {\bibinfo {author} {\bibfnamefont {D.~J.}\ \bibnamefont
  {Amit}}\ and\ \bibinfo {author} {\bibfnamefont {V.}~\bibnamefont
  {Mart\'in-Mayor}},\ }\href@noop {} {\emph {\bibinfo {title} {Field Theory,
  The Renormalization Group and Critical Phenomena}}}\ (\bibinfo  {publisher}
  {World Scientific Pub Co Inc},\ \bibinfo {year} {2005})\BibitemShut {NoStop}%
\bibitem [{\citenamefont {Brezin}\ \emph {et~al.}(1976)\citenamefont {Brezin},
  \citenamefont {\mbox{Le Guillou}},\ and\ \citenamefont
  {Zinn-Justin}}]{BrezinLeGuillouZinnJustin}%
  \BibitemOpen
  \bibfield  {author} {\bibinfo {author} {\bibfnamefont {E.}~\bibnamefont
  {Brezin}}, \bibinfo {author} {\bibfnamefont {J.~C.}\ \bibnamefont {\mbox{Le
  Guillou}}}, \ and\ \bibinfo {author} {\bibfnamefont {J.}~\bibnamefont
  {Zinn-Justin}},\ }\href@noop {} {\emph {\bibinfo {title} {Phase Transitions
  and Critical Phenomena}}}\ (\bibinfo  {publisher} {Academic Press, London,
  edited by C. Domb and M. S. A. Green, Vol. 6, p. 125},\ \bibinfo {year}
  {1976})\BibitemShut {NoStop}%
\bibitem [{\citenamefont {Carvalho}\ and\ \citenamefont
  {Sena-Junior}(2017{\natexlab{c}})}]{PhysRevD.96.116002}%
  \BibitemOpen
  \bibfield  {author} {\bibinfo {author} {\bibfnamefont {P.~R.~S.}\
  \bibnamefont {Carvalho}}\ and\ \bibinfo {author} {\bibfnamefont {M.~I.}\
  \bibnamefont {Sena-Junior}},\ }\href {\doibase 10.1103/PhysRevD.96.116002}
  {\bibfield  {journal} {\bibinfo  {journal} {Phys. Rev. D}\ }\textbf {\bibinfo
  {volume} {96}},\ \bibinfo {pages} {116002} (\bibinfo {year}
  {2017}{\natexlab{c}})}\BibitemShut {NoStop}%
\bibitem [{\citenamefont {Kleinert}\ and\ \citenamefont
  {Schulte-Frohlinde}(2001)}]{Kleinert}%
  \BibitemOpen
  \bibfield  {author} {\bibinfo {author} {\bibfnamefont {H.}~\bibnamefont
  {Kleinert}}\ and\ \bibinfo {author} {\bibfnamefont {V.}~\bibnamefont
  {Schulte-Frohlinde}},\ }\href@noop {} {\emph {\bibinfo {title} {Critical
  Properties of $\phi^{4}$ Theories}}}\ (\bibinfo  {publisher} {World
  Scientific Pub Co Inc},\ \bibinfo {year} {2001})\BibitemShut {NoStop}%
\bibitem [{\citenamefont {Bogoliubov}\ and\ \citenamefont
  {Parasyuk}(1957)}]{BogoliubovParasyuk}%
  \BibitemOpen
  \bibfield  {author} {\bibinfo {author} {\bibfnamefont {N.~N.}\ \bibnamefont
  {Bogoliubov}}\ and\ \bibinfo {author} {\bibfnamefont {O.~S.}\ \bibnamefont
  {Parasyuk}},\ }\href@noop {} {\bibfield  {journal} {\bibinfo  {journal} {Acta
  Math.}\ }\textbf {\bibinfo {volume} {97}},\ \bibinfo {pages} {227} (\bibinfo
  {year} {1957})}\BibitemShut {NoStop}%
\bibitem [{\citenamefont {Hepp}(1966)}]{Hepp}%
  \BibitemOpen
  \bibfield  {author} {\bibinfo {author} {\bibfnamefont {K.}~\bibnamefont
  {Hepp}},\ }\href@noop {} {\bibfield  {journal} {\bibinfo  {journal} {Commun.
  Math. Phys.}\ }\textbf {\bibinfo {volume} {2}},\ \bibinfo {pages} {301}
  (\bibinfo {year} {1966})}\BibitemShut {NoStop}%
\bibitem [{\citenamefont {Zimmermann}(1969)}]{Zimmermann}%
  \BibitemOpen
  \bibfield  {author} {\bibinfo {author} {\bibfnamefont {W.}~\bibnamefont
  {Zimmermann}},\ }\href@noop {} {\bibfield  {journal} {\bibinfo  {journal}
  {Commun. Math. Phys.}\ }\textbf {\bibinfo {volume} {15}},\ \bibinfo {pages}
  {208} (\bibinfo {year} {1969})}\BibitemShut {NoStop}%
\end{thebibliography}%

\end{document}